\documentstyle[12pt,fleqn]{article}
\input epsf
\def\be{\begin{equation}}
\def\ee{\end{equation}}
\def\ba{\begin{eqnarray}}
\def\ea{\end{eqnarray}}
\def\fun#1#2{\lower3.6pt\vbox{\baselineskip0pt\lineskip.9pt
        \ialign{$\mathsurround=0pt#1\hfill##\hfil$\crcr#2\crcr\sim\crcr}}}

\def\ga{\mathrel{\mathpalette\fun >}}

\begin{document}

\begin{titlepage}
\vspace*{-64pt}
\begin{flushright} {\footnotesize
FERMILAB-Pub-98/071--A\\
CERN-TH/98-59\\
OUTP-98-16-P\\
hep-ph/9802443 \\ }
\end{flushright}

\vskip 1.5cm

\begin{center}
{\Large\bf A Signature of Inflation from Dynamical Supersymmetry Breaking}

\vskip 1cm
{\bf 

 William H. Kinney$^{a,}$\footnote{E-mail: {\tt  kinneyw@fnal.gov}}
 and Antonio Riotto$^{b,}$\footnote{E-mail: {\tt riotto@nxth04.cern.ch}}$^,$\footnote{
     On leave  from Department of Theoretical Physics,
     University of Oxford, U.K. }
}
\vskip .75cm
{\it 
$^a$NASA/Fermilab Astrophysics Center \\ Fermilab
National Accelerator Laboratory, Batavia, Illinois~~60510-0500\\
\vspace{12pt}
$^b$Theory Division, CERN, CH-1211 Geneva 23, Switzerland
}
\end{center}
\vskip .5cm

\begin{quote}
In models of cosmological inflation motivated by dynamical supersymmetry breaking, the potential driving inflation may be  characterized by {\it inverse} powers of a scalar field. These models produce observables similar to those typical of the hybrid inflation scenario: negligible production of tensor (gravitational wave) modes, and a blue scalar spectral index. In this short note, we show that, unlike standard hybrid inflation models, dynamical supersymmetric inflation (DSI) predicts a measurable deviation from a power-law spectrum of fluctuations, with a variation in the scalar spectral index $|dn / d(\ln k)|$ may be as large as 0.05. DSI can be observationally distinguished from other hybrid models with cosmic microwave background measurements of the planned sensitivity of the ESA's Planck Surveyor. 
\end{quote}
\end{titlepage}

\baselineskip=24pt

\renewcommand{\baselinestretch}{1.5}

The existence of an inflationary stage during the evolution of the early
Universe is usually invoked to solve the flatness and the horizon problems of
the standard big  bang cosmology \cite{guth81}.
In the simplest picture,  the vacuum energy driving inflation is generated by
a single  scalar field $\phi$  displaced from the minimum of a 
potential $V\left(\phi\right)$. 
Quantum fluctuations of the inflaton field give rise to  temperature perturbations in the Cosmic Microwave Background (CMB) at the level of  $\delta_H=1.94\times  10^{-5}$ and these fluctuations may be responsible for the generation of structure
formation. 

The fluctuations arising from the quantum fluctuations of the inflaton field may be characterized by a power spectrum, which is the Fourier transform of the two-point density autocorrelation function. The power spectrum $\delta_H^2$  has the primordial form proportional to  $k^{n}$,  where $k$ is the amplitude of the Fourier wavevector and $n$ denotes the spectral index. 

Many  theoretically appealing models of inflation  \cite{lr} are characterized by a nearly scale invariant spectrum with the spectral index $n$  very close to unity. This is not an absolute prediction, though. Inflationary models 
generically predict that the spectral index $n$ of the density perturbation spectrum does  depend upon the scale (or the comoving wavenumber $k$), which signals the breakdown of the power-law assumption for the power spectrum \cite{bl,cop,turner}. What is even more intriguing is that some models of inflation predicting a negligible gravitational wave contribution to the microwave anisotropy, may present a strong and possibly detectable scale dependence of the spectral index. This issue has been recently analyzed in Ref. \cite{spectral} where particular attention was paid to the measurability of this scale dependence by the planned mission Planck, including the influence of polarization on the parameter estimation.  Since Planck will be able to probe a range of multipoles from $\ell=2$ to about 2000 and will measure the spectral index $n$ of a perfect power-law spectrum to an accuracy of better than $\pm 0.01$, this implies that even small variations of the spectral index with the scale,  $dn/d{\rm ln}\:k\ga 0.01$ should be measurable \cite{spectral}. (Scale dependence in the spectral index results in an increase in the measurement uncertainty by about an order of magnitude with respect to the case in which $n$ is scale independent. Other parameters are much more mildly affected.) This prediction is significant because it means that observations of the temperature anisotropies in the cosmic microwave background at the accuracy expected from Planck will allow us to distinguish between models that would otherwise be degenerate, especially in the case of models which predict no detectable gravitational wave contribution to the CMB anisotropy.

In this paper we show that in the class of supersymmetric inflationary  models dubbed  ``Dynamical Supersymmetric Inflation (DSI)'' \cite{kr},  $dn/d{\rm ln}\:k$ is detectable: if some significant scale dependence of the spectral index is measured, combined with the absence of any gravitational wave contribution to the CMB anisotropy,  this might help us in getting some deeper  insight into the nature of   supersymmetry breaking. A generic feature of models of nonperturbative gauge dynamics in supersymmetry \cite{gian} is the presence of a ``scalar field'' potential of the form
\begin{equation}
V\left(\phi\right) = {\Lambda_3^{p + 4} \over \phi^p},
\end{equation}
where the field $\phi$ is in general a label for a condensate. An inflationary phase in the very early universe is a generic and natural characteristic of dynamical supersymmetry breaking models \cite{kr}.  Like models of hybrid inflation \cite{linde91,linde94}, these models are characterized by a potential dominated by a constant term $V_0$, and require coupling to another sector to end inflation when $\phi$ reaches a critical value $\phi_c$. Unlike standard hybrid inflation models, models of this type postulate a field far from the minimum of the potential.
Models of cosmological inflation based on dynamical supersymmetry breaking are not only well motivated from a particle physics standpoint, but also very naturally meet constraints from observations of the CMB. These models have the unusual characteristic of possessing an upper limit on the total amount of inflation, as well as the feature of predicting a ``blue'' spectral index, $n > 1$. Furthermore, significantly blue spectra can occur for reasonable values of the parameters of the theory. For those values, a strong scale dependence of the spectral index is present providing us with a unique tool for confirming or disproving this class of models. 

The expressions for the spectra from slow-roll inflation are well known \cite{ll}. The slow-roll approximation during inflation  is consistent if both the slope and curvature of
the inflaton potential $V(\phi)$ are small. This condition is conventionally
expressed in terms of the ``slow-roll parameters'' $\epsilon$ and $\eta$, where
\begin{equation}
\epsilon \equiv {M_{\rm Pl}^2 \over 4 \pi} \left({H^\prime\left(\phi\right) \over H\left(\phi\right)}\right)^2 \simeq {M_{\rm Pl}^2 \over 16 \pi}
\left({V'\left(\phi\right) \over V\left(\phi\right)}\right)^2,
\end{equation}
and
\begin{equation}
\eta\left(\phi\right) \equiv {M_{\rm Pl}^2 \over 4 \pi} \left({H^{\prime\prime}\left(\phi\right) \over H\left(\phi\right)}\right) \simeq {M_{\rm Pl}^2 \over 8 \pi}
\left[{V''\left(\phi\right) \over V\left(\phi\right)} - {1 \over 2}
\left({V'\left(\phi\right) \over V\left(\phi\right)}\right)^2\right].
\end{equation}
Here $M_{\rm Pl}\simeq 1.2\times 10^{19}$ geV is the Planck mass. 
Slow-roll is then a consistent approximation for $\epsilon,\ \eta \ll 1$.  Scalar fluctuations can be
quantitatively characterized by perturbations $P_{\cal R}$ in the intrinsic
curvature
\begin{equation}
P_{\cal R}^{1/2}\left(k\right) = {1 \over \sqrt{\pi}} {H \over M_{\rm Pl}
\sqrt{\epsilon}}\Biggr|_{k^{-1} = d_H},
\end{equation}
where $H(\phi)\simeq(8\pi/3 M_{\rm Pl}^2)V(\phi)$ is the Hubble parameter during inflation. The fluctuation power is in general a function of wavenumber $k$, and is
evaluated when a given mode crosses outside the horizon during inflation,
$k^{-1} = d_H$. Outside the horizon, modes do not evolve, so the amplitude of
the mode when it crosses back {\em inside} the horizon during a later radiation
or matter dominated epoch is just its value when it left the horizon during
inflation. The  spectral index $n$ is defined by assuming an
approximately power-law form for $P_{\cal R}$ with
\begin{equation}
n - 1 \equiv {d\ln\left(P_{\cal R}\right) \over d\ln\:k},
\end{equation}
so that a scale-invariant spectrum, in which modes have constant amplitude at
horizon crossing, is characterized by $n = 1$. 

The spectral index can be expressed in terms of the slow-roll parameters
\begin{equation}
n-1= -4\epsilon +2\eta,
\end{equation}
and its derivative is given by \cite{turner}
\begin{equation}
\label{derivative}
\frac{d n}{d{\rm ln}\:k}=-8\:\epsilon^2 + 10\: \epsilon\eta -2 \xi^2,
\end{equation}
where
\begin{eqnarray}
\xi^2\equiv&& \frac{M_{\rm Pl}^4}{16 \pi^2} \frac{H^{\prime}\left(\phi\right) H^{\prime\prime\prime}\left(\phi\right)}{H^2\left(\phi\right)}\cr
\simeq&& \frac{M_{\rm Pl}^4}{64 \pi^2}\left[\frac{V^\prime\left(\phi\right) V^{\prime\prime\prime}\left(\phi\right)}{\left[V\left(\phi\right)\right]^2} - \frac{3}{2} \left({V^\prime\left(\phi\right) \over V\left(\phi\right)}\right)^2 \left({V^{\prime\prime}\left(\phi\right) \over V\left(\phi\right)}\right) + {3 \over 4} \left({V^\prime\left(\phi\right) \over V\left(\phi\right)}\right)^4\right]
\end{eqnarray}
is another slow-roll parameter. Another crucial inflationary parameter
is the influence of gravitational waves relative to density perturbation, on large-angle microwave background anisotropies, defined as a ratio of quadrupoles,
\begin{equation}
r \equiv {C_2^{\rm Tensor} \over C_2^{\rm Scalar}} \simeq 14\:\epsilon.
\end{equation}
Using this parameter, we can express (\ref{derivative}) in the following form \cite{turner}
\begin{equation}
\label{d1}
\frac{d n}{d{\rm ln}\:k}\simeq 0.12\: r^2 - 0.57\:r(1-n)- {M_{\rm Pl}^4 \over 32 \pi^2} \left({V^\prime V^{\prime\prime\prime} \over V^2}\right).\label{eqdnr}
\end{equation}
In Ref. \cite{cop} Copeland {\it et al.} concentrate on the case in which the last term in this expression, proportional to $V^{\prime\prime\prime}$, vanishes. In this case, measurable variation in $n$ is dependent on a sufficiently large tensor amplitude $r$. The situation here is exactly the opposite: DSI models are characterized by extremely small values of $r$, but nonetheless the scale dependence of the spectral index may be detected even though no trace of gravitational waves is found in the CMB anisotropy. This is because the last term in the expression (10), $\xi^2$, turns out to be crucial in DSI models and gives a large contribution to $|dn / d(\ln k)|$ during slow roll. 

We now briefly discuss the  basic features of DSI models. (The interested reader is referred to ref. \cite{kr} for more details.) The potential is described by a single degree of freedom $\phi$, of
the general form
\begin{equation}
V\left(\phi\right) = V_0 + {\Lambda_3^{p + 4} \over \phi^p}+\cdots,
\end{equation}
where the dots mean the presence of nonrenormalizable terms which are not relevant for the present discussion, but will induce a vacuum expectation  value (VEV) $\langle \phi \rangle$. 
In the
limit $\phi \ll \langle \phi\rangle$, the $\phi^{-p}$ term dominates the dynamics,
\begin{eqnarray}
V\left(\phi\right) &&\simeq  V_0 + {\Lambda_3^{p + 4} \over \phi^p},\qquad \phi
\ll \langle\phi\rangle\cr
&&= V_0 \left[1 + \alpha \left(M_{\rm Pl} \over \phi\right)^{p}\right],
\end{eqnarray}
where
\begin{equation}
\alpha \equiv {\Lambda_3^{p + 4} \over M_{\rm Pl}^p V_0}.
\end{equation}
We assume that the constant $V_0$ dominates the potential, or $\alpha \ll
\left(\phi / M_{\rm Pl}\right)^p$. (The case where the potential is dominated by the $\phi^{-p}$ term was considered in Ref. \cite{barrow93}.) In this limit, the first slow-roll parameter is
\begin{equation}
\epsilon\left(\phi\right) = {M_{\rm Pl}^2 \over 16 \pi} \left({V'\left(\phi\right)
\over V\left(\phi\right)}\right)^2 = \left({\phi_0 \over \phi}\right)^{2
\left(p + 1\right)},\label{eqepsilon}
\end{equation}
where
\begin{equation}
\left({\phi_0 \over M_{\rm Pl}}\right) = \left({p \over 4 \sqrt{\pi}}
\alpha\right)^{1/\left(p + 1\right)}.
\end{equation}
The second slow-roll parameter $\eta$ is
\begin{eqnarray}
\eta\left(\phi\right) &&= \epsilon - {M_{\rm Pl} \over 4 \sqrt{\pi}} {\epsilon'
\over \sqrt{\epsilon}}\cr
&&= \left({\phi_0 \over \phi}\right)^{2 \left(p + 1\right)} + \left({p + 1
\over 2 \sqrt{\pi}}\right) \left({\phi_0 \over \phi}\right)^{p + 1}
\left({M_{\rm Pl} \over \phi}\right).\label{eqeta}
\end{eqnarray}
Note that for $\phi \simeq \phi_0 \ll M_{\rm Pl}$, the parameter $\eta$ becomes
large, indicating a breakdown of the slow-roll approximation. In particular, it
is inconsistent to say that inflation {\em begins} at $\phi = \phi_0$, when
$\epsilon\left(\phi\right)$ in Eq. (\ref{eqepsilon}) is equal to unity, since
that expression depends on the assumption of slow-roll. However, for $\phi \gg
\phi_0$, both $\epsilon$ and $\eta$ are small and slow-roll is a consistent
approximation. In the region $\phi_0 \ll \phi \ll M_{\rm Pl}$, the second term in
(\ref{eqeta}) dominates, which is equivalent to $\eta \gg \epsilon$, and $\eta$
can be written in the useful forms
\begin{eqnarray}
\eta\left(\phi\right) &&\simeq {p + 1 \over 2 \sqrt{\pi}}
\sqrt{\epsilon\left(\phi\right)} \left({M_{\rm Pl} \over \phi}\right)\cr
&&= {p \left(p + 1\right) \over 8 \pi} \alpha \left(M_{\rm Pl} \over \phi\right)^{p
+ 2}.\label{eqformsofeta}
\end{eqnarray}
The number of e-folds $N$ is given by
\begin{eqnarray}
N = {2 \sqrt{\pi}\over M_{\rm Pl}} \int_{\phi}^{\phi_c}{d\,\phi' \over
\sqrt{\epsilon\left(\phi'\right)}} &&= \left({p + 1 \over p + 2}\right) {1
\over \eta - \epsilon}\Bigg|_\phi^{\phi_c}\cr
&&\simeq \left({p + 1 \over p + 2}\right) \left({1 \over
\eta\left(\phi_c\right)} - {1 \over \eta\left(\phi\right)}\right),\quad
\epsilon \ll \eta,
\end{eqnarray}
where $\phi_c$ is the critical value at which inflation ends. The value of
$\phi_c$ is in general determined by a coupling of the field $\phi$ to some
other sector of the theory which we have here left unspecified. Accordingly, we
will treat $\phi_c$ as simply a free parameter. Noting from Eq.\
(\ref{eqformsofeta}) that $\eta \propto \phi^{-\left(p + 2\right)}$, for $\phi
\ll \phi_c$ the number of e-folds $N$ approaches a constant, which we call
$N_{\rm tot}$,
\begin{equation}
N_{{\rm tot}} \equiv  \left({p + 1 \over p + 2}\right) {1 \over
\eta\left(\phi_c\right)} = {8 \pi \over p \left(p + 2\right)} \alpha^{-1}
\left({\phi_c \over M_{\rm Pl}}\right)^{p + 2}.
\end{equation}
This is quite an unusual feature. The total amount of inflation is bounded from above, although that upper bound can in principle be very large.
Defining $\phi_N$ to be the field value $N$ e-folds before the end of
inflation, we can then write $\eta\left(\phi_N\right)$ in terms of $N$ and
$N_{\rm tot}$ as
\begin{equation}
\eta\left(\phi_N\right) = \left({p + 1 \over p + 2}\right) {1 \over N_{\rm tot}
 - N },
\end{equation}
so that $\eta$ approaches a constant value for $N \ll N_{\rm tot}$. The spectral index is given by
\begin{eqnarray}
n - 1 &&\equiv {d\,\log\left(P_{\cal R}\right) \over
d\,\ln\:k} =  - 4 \epsilon + 2 \eta\cr
&&\simeq \left({p + 1 \over p + 2}\right) {2 \over N_{\rm tot} \left(1 - 50 /
N_{\rm tot}\right)}.
\end{eqnarray}
As announced, the spectrum turns out to be blue, but  for $N_{\rm tot} \gg 50$
the spectrum approaches scale-invariance, $n \simeq 1$. If we take the
example case of $p = 2$ and $\phi_c \sim V_0^{1/4}$, the COBE constraint on
$P_{\cal R}$ is met for $V_0^{1/4} \simeq 10^{10}\ {\rm GeV}$ and $\Lambda_3
\simeq 10^{6}\ {\rm GeV}$, very natural values for the fundamental scales in
the theory. Since $V_0^{1/4} \ll M_{\rm Pl}$, tensor modes produced during inflation are of negligible amplitude, a typical feature of hybrid inflation models. 

A blue spectrum and negligible tensor amplitude are familiar features of the great majority of hybrid inflation models. However, hybrid models typically predict no measurable scale dependence in $n$, that is, a nearly exact power-law spectrum for the density fluctuations. In DSI models, however, scale dependence of the spectral index is large for a spectrum which is significantly blue on large scales. Using now the rule $d/d{\rm ln}\:k = - d/dN$ \cite{turner}, we can express the scale dependence of the spectral index as a function of the spectral index itself
\begin{equation}
\frac{dn}{d{\rm ln}\:k}= - \frac{1}{2}\left({p + 2 \over p + 1}\right)(n-1)^2.
\end{equation}
Taking $p = 2$ and $n = 1.2$, we have $d n / d\left(\ln k\right) \simeq - 0.05$, a value readily measurable by Planck \cite{cop}. Note that in all cases, $n$ approaches the scale-invariant limit $n = {\rm const.} \simeq 1$ on small scales, which serves to mitigate difficulties which a strongly blue spectrum would otherwise have meeting constraints from structure formation or from production of primordial black holes \cite{carr94,green97}. Figure 1 shows the scalar spectral index as a function of e-folds for several choices of parameters.

In conclusion, we have shown that the class of models of inflation inspired by the idea of dynamical supersymmetry breaking may be be characterized by a significant and measurable deviation from the power-law spectrum of fluctuations. This opens up the possibility of distinguishing this class of models from the standard hybrid inflation models and, if the predicted signature is found, of  looking at dynamical supersymmetry breaking on the sky.

\vspace{36pt}
\centerline{\bf ACKNOWLEDGMENTS}
\vspace{24pt}
We would like to thank A.R. Liddle for
discussions.  WHK would like to thank the CERN Theory Group where part of this work was done, for the kind hospitality. WHK is  supported by the DOE and NASA under
Grant NAG5-2788.

\begin{figure}
\centerline{\epsfxsize=300pt \epsfbox{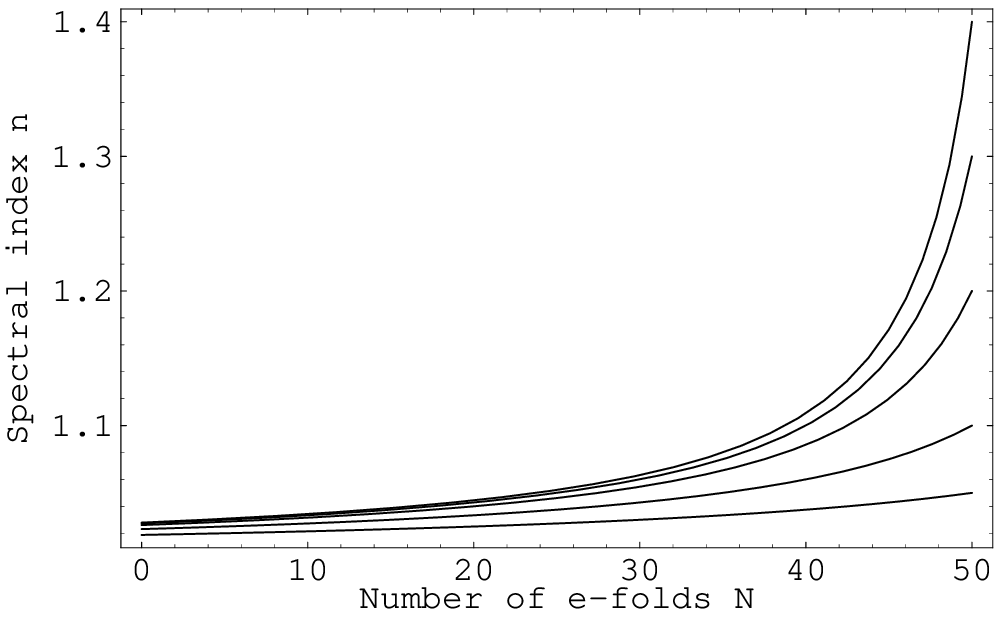}}
\hspace*{1em} {\footnotesize{{\bf Fig.\ 1:} Scalar spectral index $n$ as a function of the number of e-folds $N \propto \ln(k)$ for several choices of parameters. Note especially the rapid approach to scale-invariance at short wavelengths (small $N$).}}
\end{figure}

\frenchspacing
\def\prpts#1#2#3{Phys. Reports {\bf #1}, #2 (#3)}
\def\prl#1#2#3{Phys. Rev. Lett. {\bf #1}, #2 (#3)}
\def\prd#1#2#3{Phys. Rev. D {\bf #1}, #2 (#3)}
\def\plb#1#2#3{Phys. Lett. {\bf #1B}, #2 (#3)}
\def\npb#1#2#3{Nucl. Phys. {\bf B#1}, #2 (#3)}
\def\apj#1#2#3{Astrophys. J. {\bf #1}, #2 (#3)}
\def\apjl#1#2#3{Astrophys. J. Lett. {\bf #1}, #2 (#3)}
\begin{picture}(400,50)(0,0)
\put (50,0){\line(350,0){300}}
\end{picture}

\end{document}